\documentstyle[multicol,aps,prl,twocolumn,epsf]{revtex}
\newcommand{\bm}[1]{\mbox{\boldmath $#1$}}

\begin{document}
\draft 
\topmargin=0.01cm
\twocolumn[\hsize\textwidth\columnwidth\hsize\csname@twocolumnfalse\endcsname
\title{Inverse cascade in two-dimensional turbulence\,: deviations
from Gaussianity}
\author{G.~Boffetta$^{1}$, A.~Celani$^{1,2}$ and M.~Vergassola$^{2}$ \\
\small{$^{1}$ Dipartimento di Fisica Generale, Universit\`a di Torino,
and INFM Unit\`a di Torino Universit\`a, 
I--10126  Torino, Italy.}\\
\small{$^{2}$ CNRS, Observatoire de la C\^ote d'Azur, B.P. 4229,
06304 Nice Cedex 4, France.}}
\date{\today}
\maketitle
\begin{abstract}
High resolution numerical simulations of stationary inverse energy
cascade in two-dimensional turbulence are presented. Deviations from
Gaussianity of velocity differences statistics are quantitatively
investigated. The level of statistical convergence is pushed enough to
permit reliable measurement of the asymmetries in the probability
distribution functions of longitudinal increments and odd-order
moments, which bring the signature of the inverse energy flux. Their
scaling laws do not present any measurable intermittency
correction. The seventh order skewness is found to increase by almost
two orders of magnitude with respect to the third, thus becoming of
order unity.

\end{abstract}
\pacs{PACS number(s)\,: 47.10.+g, 47.27.-i, 05.40.+j}]

The inverse energy cascade in two dimensional Navier-Stokes turbulence
is one of the most important phenomena in fluid dynamics. In agreement
with the remarkable prediction of R.H.~Kraichnan in 1967 \cite{RHK67},
the coupled constraints of energy and enstrophy conservation make the
energy injected into the system flow toward the large scales. This is 
a basic difference with respect to 3D turbulence, where energy flows
toward small scales in a direct cascade. The dynamical process of
structuring and organization of the large scales by the inverse
cascade is also of great interest for geophysical fluid
dynamics. First numerical and experimental observations of the inverse
cascade and the ensuing Kolmogorov energy spectrum were obtained in 
\cite{Lilly72,FMJ77,AS81,HMM83,FS84,HMcW85,JSom,MV}. The important point
foreseen in \cite{AS81} is that the smallness of the skewness suggests
that intermittency might be weak. This conjecture was later supported
by numerical simulations \cite{SY94,DBPT99} and experiments
\cite{PT98}\,: scaling laws are compatible with dimensional
predictions and both transversal and longitudinal velocity probability
distribution functions (pdf's) look not far from Gaussian. The
evidence stemming from experiments and simulations is that the inverse
transfer takes place via clustering of small-scale equal sign
vortices. Strong deviations from Gaussianity appear if the system has
a finite size and friction extracting the energy from the large scales
is small (or absent). A pile-up of energy akin to the Bose-Einstein
condensation takes then place in the gravest mode \cite{RHK67}, large
scale vortices are formed and energy spectra steeper than the
Kolmogorov one are observed \cite{VBor}. Here we shall not consider
the condensation phase, concentrating on the inverse cascade
statistics. Theoretically, inverse cascade enjoys a great advantage
with respect to the direct one\,: the limit of molecular viscosity
$\nu\to 0$ can be taken without any harm in the equations of motion
for velocity structure functions. At variance of 3D turbulence, the
energy dissipation $\nu \langle\left(\nabla{\bm v}\right)^2\rangle$ is
indeed vanishing when $\nu\to 0$. The absence of dissipative anomalies
is the clue for the analytical solution of inverse cascades in passive
scalar advection \cite{CKV,GV}. Intermittency was found to be absent,
even though the statistics might be strongly differing from
Gaussian. For 2D Navier-Stokes inverse cascade, dissipative terms can
again be discarded but the situation is complicated by pressure
gradients.  They couple indeed the statistics of velocity differences
$\delta_r {\bm v}\equiv {\bm v}({\bm r})-{\bm v}({\bm 0})$ at various
${\bm r}$'s in a non-local way. Closures on velocity
increments-pressure gradients correlations have been proposed by
invoking the quasi-Gaussianity of the statistics and quantitative
predictions have been derived in this way \cite{VY99}.  The issue of
quasi-Gaussianity is however moot as deviations are intrinsically
entangled to the dynamical process of inverse energy cascade.
Standard calculations (see, e.g., \cite{UF95}) permit indeed to derive
the $3/2$ Kolmogorov law for 2D turbulence\,: $S^{(3)}_{L}(r)=\langle
\left[\delta_r {\bm v} \cdot\hat{\bm r}\right]^3\rangle=3/2\epsilon
r$, where $ \hat{\bm r}={\bm r}/r$.  The energy flux is denoted by
$\epsilon$ and the fact that it goes upscale reflects into the
positive sign of the moment.  Precise quantitative informations on the
deviations from Gaussianity are however difficult to obtain. Odd-order
structure functions involve for example strong cancellations between
negative and positive contributions and the $3/2$ law itself could not
be observed in previous studies, due to lack of resolution and/or
statistical convergence. It is our purpose here to present the results
of high-resolution numerical simulations aimed at quantitatively
analyzing deviations from Gaussianity in the inverse energy cascade.

Specifically, the 2D Navier-Stokes equation for the vorticity 
$\omega({\bm r},t)=-\Delta\psi({\bm r},t)$ is\,:
\begin{equation}
\label{NS2D}
\partial_t\omega+J\left(\omega,\psi\right)=\nu\Delta\omega
-\alpha\omega-\Delta f,
\end{equation}
where $\psi$ is the stream function, the velocity ${\bm
v}=\nabla^{\perp}\psi=(\nabla_y\psi,-\nabla_x\psi)$ and $J$ denotes
the Jacobian. The friction linear term $-\alpha \omega$ extracts
energy from the system at scales comparable to the friction scale
$\eta_{\rm fr}\sim \epsilon^{1/2} \alpha^{-3/2}$, assuming a
Kolmogorov scaling law for the velocity. To avoid Bose-Einstein
condensation in the gravest mode we choose $\alpha$ to make $\eta_{\rm
fr}$ sufficiently smaller than the box size. The other relevant length
in the problem is the small-scale forcing correlation length $l_f$,
bounding the inertial range for the inverse cascade as $l_f \ll r\ll
\eta_{\rm fr}$. We use a Gaussian forcing with correlation function
$\langle f({\bm r},t)\,f({\bm 0},t')
\rangle=\delta(t-t')\,F(r/l_f)$. The $\delta$-correlation in time
ensures the exact control of the energy injection rate.  The forcing
space correlation should decay rapidly for $r\gg l_f$ and we choose
$F(x)= F_0 l_f^2 \exp(-x^2/2)$, where $F_0$ is the energy input.  The
numerical integration of (\ref{NS2D}) is performed by a standard
$2/3$-dealiased pseudospectral method on a doubly periodic square
domain of $N^2\!=\!2048^2$ grid points.  The viscous term in
(\ref{NS2D}) has the role of removing enstrophy at scales smaller
than $l_f$ and, as customary, it is numerically more convenient to
substitute it by a hyperviscous term (of order eight in our
simulations).  Time evolution is obtained by a standard second-order
Adams-Bashforth scheme.  After the system has reached stationarity,
analysis is performed over twenty snapshots of the velocity field
equally spaced by one large-eddy turnover time.

\begin{figure}
\epsfxsize=8truecm
\epsfbox{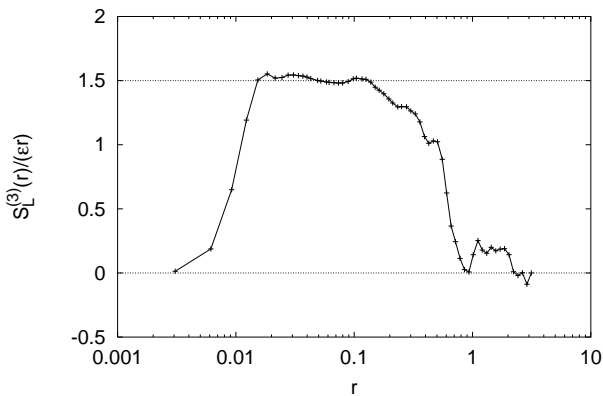}
\caption{Compensated third order longitudinal structure function 
$S^{(3)}_{L}(r)/(\epsilon r)$. The dotted line is the value $3/2$. 
Note the linear vertical scale.}
\label{f1}
\end{figure}

\begin{figure}
\epsfxsize=8truecm
\epsfbox{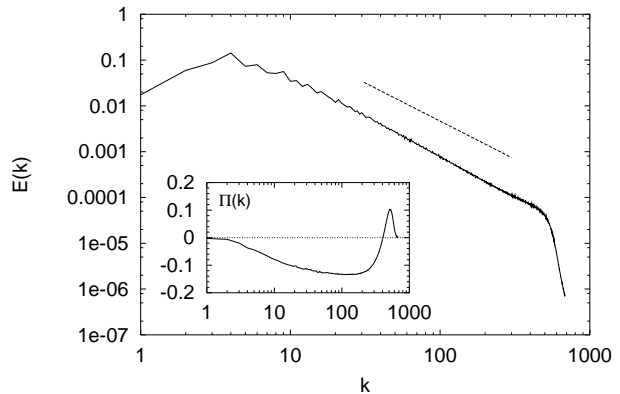}
\caption{Energy spectrum $E(k)$. The dashed line has a 
Kolmogorov power law scaling.  In the inset is shown the energy flux $\Pi(k)$.}
\label{f2}
\end{figure}

\begin{figure}
\epsfxsize=8truecm
\epsfbox{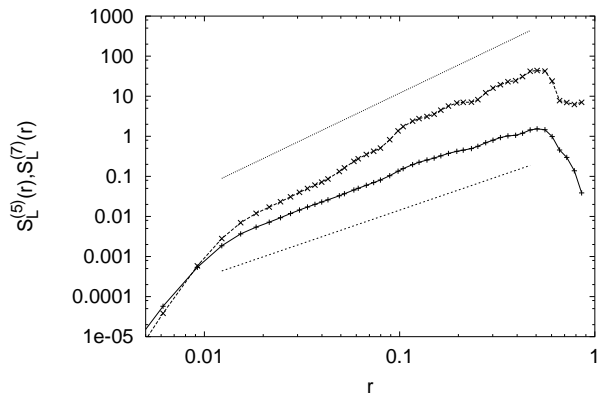}
\caption{Longitudinal structure functions of order $5$ (lower) and $7$ (upper).
Power laws with exponents $5/3$ and $7/3$ are drawn for comparison.}
\label{f3}
\end{figure}

Let us now discuss the results. In Fig.~\ref{f1} we present the
third-order longitudinal structure function $S^{(3)}_{L}(r)$
compensated by the factor $1/(\epsilon r)$, showing a neat plateau at
the value $3/2$ -- in agreement with the Kolmogorov law -- over a
range of almost one decade of scales. In Fig.~\ref{f2} it is presented the
energy spectrum $E(k)$, which displays a clear Kolmogorov scaling
$k^{-5/3}$, and the energy flux $\Pi(k)$.
\narrowtext
The Kolmogorov constant in
\begin{equation}
\label{CK}
E(k)=C\,\epsilon^{2/3}\,k^{-5/3},
\end{equation}
is found to be $C=6.0\pm0.4$ .  Previous numerical simulations and
experiments report values of the Kolmogorov constant $C$ ranging from
$5.8$ to $7.0$ \cite{FS84,JSom,MV,SY94,DBPT99,PT98,VBor}.  The
structure function constants corresponding to (\ref{CK}) are
$C^{(2)}_{L}=3C^{(2)}_{T}/5=\frac{\sqrt{3} \pi}{2^{5/3} \Gamma(4/3)^2}
C = 12.9\pm0.8$, where the first two equalities follow from isotropy
and incompressibility and
\begin{equation}
\label{S2}
S^{(n)}_{L}(r)= \langle \left[\delta_r{\bm v}\cdot\hat{\bm
r}\right]^n\rangle=C^{(n)}_{L}\left(\epsilon\,r\right)^{n/3}\,.
\end{equation}
\narrowtext
For transverse moments, $\hat{\bm r}$ is substituted in (\ref{S2}) by
$\hat{\bm r}_{\perp}$, perpendicular to it.  It is of interest to
remark that longitudinal and transverse velocity increments are
uncorrelated, i.e. $S^{(2)}_{L,T}(r)=0$.  The relatively large value
of $C^{(2)}_L$ implies a small skewness of the longitudinal velocity
differences $(3/2)/(C^{(2)}_{L})^{3/2}= 0.03$.  Albeit the
longitudinal pdf looks close to Gaussian and quite symmetric,
nevertheless on a more quantitative ground asymmetries turn out to be
quite strong as shown by the two curves of $S^{(5)}_{L}(r)$ and
$S^{(7)}_{L}(r)$ in Fig.~\ref{f3}.
\narrowtext
First, we can observe that their scaling behavior is in agreement with
Kolmogorov predictions, without significant anomalous
corrections. Second, the constants are $C^{(5)}_{L}\simeq 130$ and
$C^{(7)}_{L}\simeq 14000$, giving for the hyper-skewness
$C^{(5)}_{L}/(C^{(2)}_{L})^{5/2} \simeq 0.22$ and
$C^{(7)}_{L}/(C^{(2)}_{L})^{7/2} \simeq 1.8$.  The error bars can be
estimated from r.m.s. fluctuations of compensated plots and for the
seventh order (which is of course the most delicate) they amount to
$20\%$.  The increase of the skewness by almost two orders of
magnitude from the third to the seventh order is particularly
informative. H\"older inequalities apply indeed to absolute moments
and odd-order moments (without absolute values as in (\ref{S2})) might
{\it a priori} even become smaller when their order increases.  Our
main motivation was precisely to find out whether odd-order moments
were decreasing or increasing with the order. The answer to this
question shows that hyperskewness is definitely not a ``small
parameter'' to be used in perturbative schemes for the statistical
properties of the inverse energy cascade. The predictions in
\cite{VY99}, although based on a closure explictly invoking small
deviations from Gaussianity, turn out to be compatible with the
numerical results.  This indicates that the closure is likely to be
more robust and ``nonperturbative'' than its derivation might suggest.

\begin{figure}
\epsfxsize=8truecm
\epsfbox{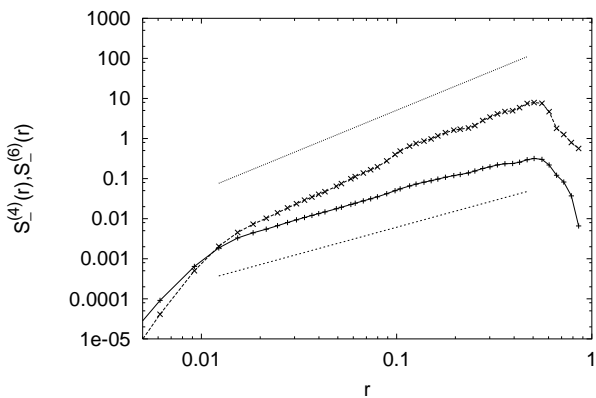}
\caption{Antisymmetric structure functions $S^{(n)}_{-}(r)$ of order $n=4,6$.
The solid lines have slopes $4/3$ and $2$, respectively.}
\label{f4}
\end{figure}

Another striking evidence for the importance of the longitudinal pdf
asymmetries is provided in Fig.~\ref{f4}. 
\narrowtext
\narrowtext
\begin{figure}
\epsfxsize=8truecm
\epsfbox{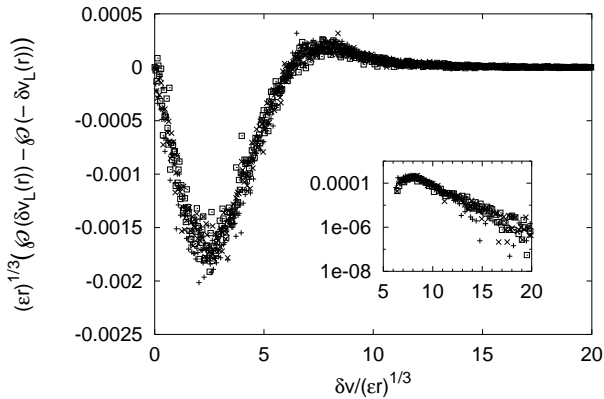}
\caption{Antisymmetric part of the longitudinal velocity
increments pdf, rescaled according to Kolmogorov scaling, at separations
$r=0.05$ ($\Box$), $r=0.075$ ($\times$), $r=0.1$ ($+$), 
lying in the inertial range of scales.
In the inset, the same data in 
linear-logarithmic scale.}
\label{f4bis}
\end{figure}
We consider here the
antisymmetric part of the pdf ${\cal P}\left(\delta
v_L(r)\right)-{\cal P}\left(-\delta v_L(r)\right)$ (shown in
Fig.~\ref{f4bis}) and calculate ``antisymmetric structure functions'' such as
$S^{(4)}_{-}=\int_0^{\infty} u^4\left( {\cal P}(u)-{\cal
P}(-u)\right)\,du$.  
Both the fourth and the sixth moment have a clean
Kolmogorov scaling $S^{(n)}_{-}(r)= C^{(n)}_{-} \left( \epsilon r
\right)^{n/3}$, with $C^{(4)}_{-}/(C^{(2)}_L)^{2}\simeq 0.08$ and
$C^{(6)}_{-}/(C^{(2)}_L)^{3} \simeq 0.6$. This indicates that the
non-Gaussian antisymmetric part, although visually small, has
imprinted all the relevant scaling informations on the inverse cascade.

  From the graph in the inset of Fig.~\ref{f4bis} it can be appreciated
that the asymmetric part of the pdf has relatively consistent tails, which
are the cause for the large observed values of the hyperskewness of
order $5$ and $7$.

\begin{figure}
\epsfxsize=8truecm
\epsfbox{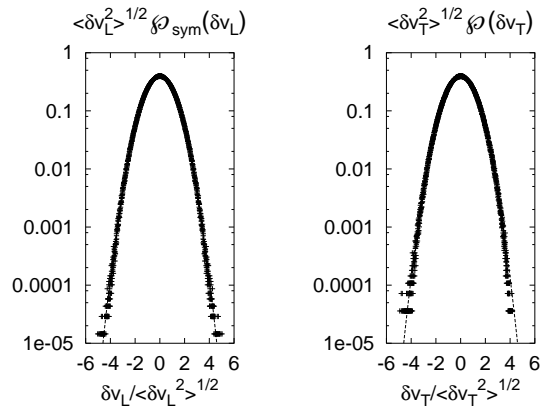}
\caption{Left: symmetric part of the longitudinal velocity
difference pdf. Right: pdf of transverse velocity differences.
The forcing is restricted to a band of wavenumbers and the data refer
to a single velocity field snapshot. Gaussian distributions are shown 
as solid lines.}
\label{f5}
\end{figure}

To investigate possible dependencies on the type of forcing we also
performed (shorter) numerical simulations with an injection rate
characterized by the spectral correlation function $\langle f({\bm
k},t)\,f({\bm k}',t') \rangle=\delta(t-t') \delta({\bm k}+{\bm k}')
\delta(1-kl_f)$ \cite{FS84}. At variance with the former choice, this
forcing is limited to a narrow bandwidth in Fourier space but its
spatial correlations decay rather slowly.  Odd-order structure
functions and the antisymmetric part of the pdf do not show any
visible dependence on the details of the energy input. Conversely, the
symmetric part of the pdf of velocity differences and even order
moments are more sensitive. For the forcing limited to a shell of
wavenumbers, both the (symmetrized) longitudinal and the transverse
pdfs are visually indistinguishable from Gaussian, as shown in
Fig.~\ref{f5}.
\narrowtext
Deviations
of kurtosis and hyperkurtosis from their Gaussian values are small and
compatible with those presented in ~\cite{PT98}. For the forcing
localized in physical space, the far tails of the pdf at scales
$O(l_f)$ tend to be broader.  This tendency is due to the formation of
small vortices of size comparable to $l_f$, which generate large
velocity differences -- especially transverse ones -- across a
distance of the order of their size. The effect becomes of course
negligible at scales larger than $l_f$ but it might affect the quality
and the extension of the scaling region for even order structure
functions \cite{BPPSV86}.

In conclusion, we have presented quantitative evidences for deviations
from Gaussianity of the velocity increment statistics in the inverse
energy cascade. Odd-order structure functions display a clean power
law scaling compatible with classical Kolmogorov predictions.
Numerical prefactors in adimensionalized structure functions are
expected to be universal with respect to the forcing statistics and
have been measured up to the seventh order. Despite the small value of
the skewness, deviations have been shown to be quite strong and the
hyperskewness of seventh order to be of order unity.  Asymmetries in
longitudinal velocity statistics should therefore be incorporated and
treated systematically in theoretical models for the inverse energy
cascade.

\medskip
{\bf Acknowledgements.} We are grateful to A.~Babiano, G.~Falkovich,
K.~Gaw\c{e}dzki, A.~Mazzino, A.~Pouquet, P.~Tabeling and V.~Yakhot for useful
discussions.  Support from the ESF-TAO programme (AC), from the
network "Intermittency in Turbulent systems" under contract
FMRX-CT98-0175 and from INFM "PRA TURBO" (AC and GB), is gratefully
acknowledged.  Numerical simulations were performed at IDRIS, under
the contract No. 991226, and partially at CINECA within the project
"Lagrangian and Eulerian statistics in fully developed turbulence".


\begin{thebibliography}{99}
\bibitem{RHK67} 
R.H.~Kraichnan, {\it Phys. Fluids}, {\bf 10}, 1417, (1967).
\bibitem{Lilly72}
D.~K.~Lilly, {\it Geophys. Fluid Dyn.},{\bf 3}, 290, (1972). 
\bibitem{FMJ77}
D.~Fyfe, D.~Montgomery \& G.~Joyce, {\it J. Plasma Physics},{\bf 17}, 
369, (1977).
\bibitem{AS81}
E.~Siggia \& H.~Aref, {\it Phys. Fluids}, {\bf 24}, 171, (1981).
\bibitem{HMM83}
M.~Hossain, W.~H.~Matthaeus \& D.~Montgomery, 
{\it J. Plasma Physics}, {\bf 30}, 479, (1983). 
\bibitem{FS84}
U.~Frisch \& P.L.~Sulem, {\it Phys. Fluids}, {\bf 27}, 1911, (1984). 
\bibitem{HMcW85}
J.~R.~Herring \& J.~C.~McWilliams, {\it J. Fluid. Mech},{\bf 153}, 229, (1985).
\bibitem{JSom}
J.~Sommeria, {\it J. Fluid Mech.}, {\bf 170}, 139, (1986). 
\bibitem{MV}
M.~E.~Maltrud, \& G.~K.~Vallis, {\it J. Fluid Mech.}, {\bf 228}, 321, (1991).
\bibitem{SY94}
L.~Smith \& V.~Yakhot, {\it Phys. Rev. Lett.}, {\bf 71}, 352, (1993).
\bibitem{DBPT99}
T.~Dubos, A.~Babiano, J.~Paret \& P.~Tabeling, {\it Numerical investigation 
of internal intermittency in the two-dimensional inverse energy cascade},
preprint, (1999).
\bibitem{PT98}
J.~Paret \& P.~Tabeling, {\it Phys. Fluids}, {\bf 10}, 3126, (1998).
\bibitem{VBor}
V.~Borue, {\it Phys. Rev. Lett.}, {\bf 72}, 1475, (1994).
\bibitem{CKV} 
M.~Chertkov, I.~Kolokolov \& M.~Vergassola, {\it Phys. Rev. Lett.}, 
{\bf 80}, 512, (1998).
\bibitem{GV} K.~Gaw\c{e}dzki \& M.~Vergassola, chao-dyn/9811399,
(1998).
\bibitem{VY99}
V.~Yakhot, chao-dyn/9904016, (1999).
\bibitem{UF95}
U.~Frisch, {\it Turbulence}, Cambridge Univ. Press, Cambridge, (1995).
\bibitem{BPPSV86} 
The asymmetric part of the pdfs is not affected when high vorticity
regions are removed as in R.~Benzi , G.~Paladin, S.~Patarnello,
P.~Santangelo \& A.~Vulpiani, {\it J. Phys. A: Math. Gen.} {\bf 19},
3771, (1986).  This confirms the expectation that the intense small
scale coherent vortices generated by the direct enstrophy cascade play
no essential role in the energy transfer towards larger scales.

\end{thebibliography}
\end{document}